\documentclass[aps,prl,preprint,tightenlines,superscriptaddress,showpacs,byrevtex]{revtex4}
\usepackage{graphicx}% Include figure files
\usepackage{dcolumn}% Align table columns on decimal point
\usepackage{bm}% bold math

\begin{document}

\vspace*{-3\baselineskip}
\resizebox{!}{3cm}{\includegraphics{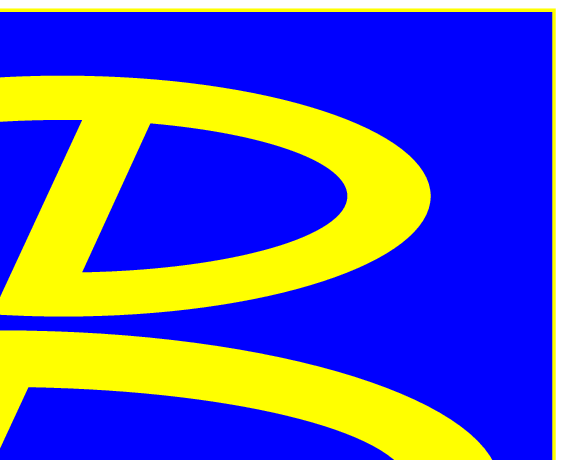}}

%\preprint{Ver.2.0}
\preprint{KEK Preprint 2002-69}
\preprint{Belle Preprint 2002-23}

%%%%%%%%%%%%%%%%%%%%%%
% New commands
%%%%%%%%%%%%%%%%%%%%%%
\newcommand{\ups}{\Upsilon (4S)}
\newcommand{\ra}{\rightarrow}
\newcommand{\myindent}{\hspace*{2cm}}  % My indent for equation environments
\newcommand{\DMDVAL}{0.494}
\newcommand{\DMDERRSTA}{0.012}
\newcommand{\DMDERRSYS}{0.015}
\newcommand{\DMDRESSTA}{\DMDVAL~\pm~\DMDERRSTA({\rm stat})~({\rm ps}^{-1})}
\newcommand{\DMDRESULT}{\DMDVAL~\pm~\DMDERRSTA({\rm stat})~\pm~\DMDERRSYS({\rm syst})~{\rm ps}^{-1}}
\newcommand{\DMDRESULTNEW}{(\DMDVAL~\pm~\DMDERRSTA~\pm~\DMDERRSYS)~{\rm ps}^{-1}}
\newcommand{\DMDRESULTNEWNEW}{(\DMDVAL~\pm~\DMDERRSTA({\rm stat})~\pm~\DMDERRSYS({\rm syst}))~{\rm ps}^{-1}}
\newcommand{\dt}{\Delta t}
\newcommand{\dz}{\Delta z}
\newcommand{\dmd}{\Delta m_d}
\newcommand{\taub}{{\tau}_{B^0}}
\newcommand{\bz}{B^0}
\newcommand{\bzb}{\overline{B}{}^0}
\newcommand{\mdiff}{M_{\rm diff}}
\newcommand{\dslnu}{D^{*-}\ell^+\nu}
\newcommand{\bzdslnu}{\bz \to \dslnu}
\newcommand{\ztag}{z_{\rm tag}}
\newcommand{\zrec}{z_{\rm rec}}
\newcommand{\dzb}{\overline{D}{}^0}
\newcommand{\kppm}{K^+\pi^-}
\newcommand{\kppmpz}{\kppm\pi^0}
\newcommand{\kppmpppm}{\kppm\pi^+\pi^-}
\newcommand{\dzbkppm}{\dzb \to \kppm}
\newcommand{\dzbkppmpz}{\dzb \to \kppmpz}
\newcommand{\dzbkppmpppm}{\dzb \to \kppmpppm}
\newcommand{\thetabdl}{\theta_{B,D^*\ell}}
\newcommand{\cosbdl}{\cos\thetabdl}
\newcommand{\mnu}{M_\nu}
\title{Measurement of the
{\boldmath $\bz$}-{\boldmath $\bzb$} Mixing Parameter {\boldmath $\dmd$}
 using Semileptonic {\boldmath $\bz$} Decays}

% Paper:    B-Bbar mixing via semileptonic B0 decays
% Journal:  Physical Review Letters
% Contacts: K. Hara (kohji@champ.hep.sci.osaka-u.ac.jp)
%           M. Hazumi (masashi.hazumi@kek.jp)
% Deadline: Thu Jul 10 2002 01:00 am JST
% Non-responding authors or those who said NO are commented out.
% ====================================================================
% Click the RELOAD button on your web browser to see the updated file.
% ====================================================================
% Use \input{author} to insert this material into your latex file.
%%%%% Force institutions to appear in alphabetical order when typeset.
\affiliation{Aomori University, Aomori}
\affiliation{Budker Institute of Nuclear Physics, Novosibirsk}
\affiliation{Chiba University, Chiba}
\affiliation{Chuo University, Tokyo}
\affiliation{University of Cincinnati, Cincinnati OH}
\affiliation{University of Frankfurt, Frankfurt}
\affiliation{Gyeongsang National University, Chinju}
\affiliation{University of Hawaii, Honolulu HI}
\affiliation{High Energy Accelerator Research Organization (KEK), Tsukuba}
\affiliation{Hiroshima Institute of Technology, Hiroshima}
\affiliation{Institute of High Energy Physics, Chinese Academy of Sciences, Beijing}
\affiliation{Institute of High Energy Physics, Vienna}
\affiliation{Institute for Theoretical and Experimental Physics, Moscow}
\affiliation{J. Stefan Institute, Ljubljana}
\affiliation{Kanagawa University, Yokohama}
\affiliation{Korea University, Seoul}
\affiliation{Kyoto University, Kyoto}
\affiliation{Kyungpook National University, Taegu}
\affiliation{Institut de Physique des Hautes \'Energies, Universit\'e de Lausanne, Lausanne}
\affiliation{University of Ljubljana, Ljubljana}
\affiliation{University of Maribor, Maribor}
\affiliation{University of Melbourne, Victoria}
\affiliation{Nagoya University, Nagoya}
\affiliation{Nara Women's University, Nara}
\affiliation{National Kaohsiung Normal University, Kaohsiung}
\affiliation{National Lien-Ho Institute of Technology, Miao Li}
\affiliation{National Taiwan University, Taipei}
\affiliation{H. Niewodniczanski Institute of Nuclear Physics, Krakow}
\affiliation{Nihon Dental College, Niigata}
\affiliation{Niigata University, Niigata}
\affiliation{Osaka City University, Osaka}
\affiliation{Osaka University, Osaka}
\affiliation{Panjab University, Chandigarh}
\affiliation{Peking University, Beijing}
\affiliation{Princeton University, Princeton NJ}
\affiliation{RIKEN BNL Research Center, Brookhaven NY}
%%%\affiliation{Saga University, Saga}
\affiliation{University of Science and Technology of China, Hefei}
\affiliation{Seoul National University, Seoul}
\affiliation{Sungkyunkwan University, Suwon}
\affiliation{University of Sydney, Sydney NSW}
\affiliation{Tata Institute of Fundamental Research, Bombay}
\affiliation{Toho University, Funabashi}
\affiliation{Tohoku Gakuin University, Tagajo}
\affiliation{Tohoku University, Sendai}
\affiliation{University of Tokyo, Tokyo}
\affiliation{Tokyo Institute of Technology, Tokyo}
\affiliation{Tokyo Metropolitan University, Tokyo}
\affiliation{Tokyo University of Agriculture and Technology, Tokyo}
\affiliation{Toyama National College of Maritime Technology, Toyama}
\affiliation{University of Tsukuba, Tsukuba}
\affiliation{Utkal University, Bhubaneswer}
\affiliation{Virginia Polytechnic Institute and State University, Blacksburg VA}
\affiliation{Yokkaichi University, Yokkaichi}
\affiliation{Yonsei University, Seoul}
  \author{K.~Hara}\affiliation{Osaka University, Osaka} % Osaka
  \author{M.~Hazumi}\affiliation{High Energy Accelerator Research Organization (KEK), Tsukuba} % KEK
  \author{K.~Abe}\affiliation{High Energy Accelerator Research Organization (KEK), Tsukuba} % KEK
  \author{K.~Abe}\affiliation{Tohoku Gakuin University, Tagajo} % TohokuGakuin
% \author{N.~Abe}\affiliation{Tokyo Institute of Technology, Tokyo} % TIT
% \author{R.~Abe}\affiliation{Niigata University, Niigata} % Niigata
  \author{T.~Abe}\affiliation{Tohoku University, Sendai} % Tohoku
  \author{I.~Adachi}\affiliation{High Energy Accelerator Research Organization (KEK), Tsukuba} % KEK
  \author{Byoung~Sup~Ahn}\affiliation{Korea University, Seoul} % Korea
  \author{H.~Aihara}\affiliation{University of Tokyo, Tokyo} % Tokyo
  \author{M.~Akatsu}\affiliation{Nagoya University, Nagoya} % Nagoya
% \author{M.~Asai}\affiliation{Hiroshima Institute of Technology, Hiroshima} % Hiroshima
  \author{Y.~Asano}\affiliation{University of Tsukuba, Tsukuba} % Tsukuba
% ...
% Tsukuba
\collaboration{The Belle Collaboration}

\date{\today}% It is always \today, today,

\begin{abstract}
We present a measurement of the $\bz$-$\bzb$ mixing parameter 
$\dmd$ using neutral $B$ meson pairs in a 29.1~fb$^{-1}$ 
data sample collected at the $\Upsilon(4S)$ resonance with the Belle 
detector at the KEKB asymmetric-energy $e^+e^-$ collider.
We exclusively reconstruct one neutral $B$ meson in the semileptonic 
$B^0 \rightarrow D^{*-}\ell^+\nu$ decay mode and identify the flavor of
the accompanying $B$ meson from its decay products. From the
distribution of the time intervals between the two 
flavor-tagged $B$ meson decay points, we obtain 
$\Delta m_d = \DMDRESULTNEW$, where the first error is statistical and
the second error is systematic.
\end{abstract}


\maketitle

%%%%%%%%%%%%%%%%
%\section{Introduction}
%%%%%%%%%%%%%%%%

$\bz$-$\bzb$ mixing plays a unique role in the determination
of basic parameters in the standard model (SM)
of elementary particles.
It is characterized by the oscillation frequency
$\dmd$, which is the difference between the
two mass eigenvalues of neutral $B$ meson states.
In the SM, the mixing is due to second-order weak interactions
known as box diagrams~\cite{Mixing} whose amplitudes 
involve $V_{td}$, an element
of the quark mixing matrix governing transitions
between the top and down quarks~\cite{CKM}.
The mixing also induces
large time-dependent $CP$ violation 
in neutral $B$ meson decays, which has been 
observed recently~\cite{CP1_Belle,CP1_BaBar}.
For such $CP$ violation measurements,
precise $\dmd$ measurements are important.

In this Letter, we report a $\dmd$ measurement with
31.3 million $\bz\bzb$ pairs, collected  with
the Belle detector at the KEKB asymmetric-energy
$e^+e^-$ (3.5 on 8~GeV) collider~\cite{KEKB}
operating at the $\Upsilon(4S)$ resonance.
The time evolution is described as
${e^{-|\dt|/\taub}}/(4\taub)\{1 \pm \cos (\dmd \Delta t)\}$, where
the plus (minus) sign is taken when the
flavor of one $B$ meson is opposite to (the same as) the other,
$\taub$ is the lifetime of the neutral
$B$ meson and $\dt$ is the proper time difference between the two $B$
meson decays.  
At KEKB, the $\Upsilon(4S)$ is produced
with a Lorentz boost of $\beta\gamma=0.425$ nearly along
the electron beamline ($z$).
Since the $B$ mesons are approximately at 
rest in the $\Upsilon(4S)$ center-of-mass system (cms),
$\Delta t$ can be determined from the displacement in $z$ 
between the two $B$ decay vertices:
$\Delta t \simeq (\zrec - \ztag)/\beta\gamma c
 \equiv \Delta z/\beta\gamma c$.

The Belle detector~\cite{Belle} is a large-solid-angle
spectrometer that
consists of a silicon vertex detector (SVD),
a central drift chamber (CDC), an array of
aerogel threshold \v{C}erenkov counters (ACC), 
time-of-flight
scintillation counters (TOF), and an electromagnetic calorimeter
comprised of CsI(Tl) crystals (ECL)  located inside 
a superconducting solenoid coil that provides a 1.5~T
magnetic field.  An iron flux-return located outside of
the coil is instrumented to detect $K_L^0$ mesons and to identify
muons (KLM).  

\begin{figure}[htbp]
  \begin{center}
    \includegraphics[width=0.6\textwidth,clip]{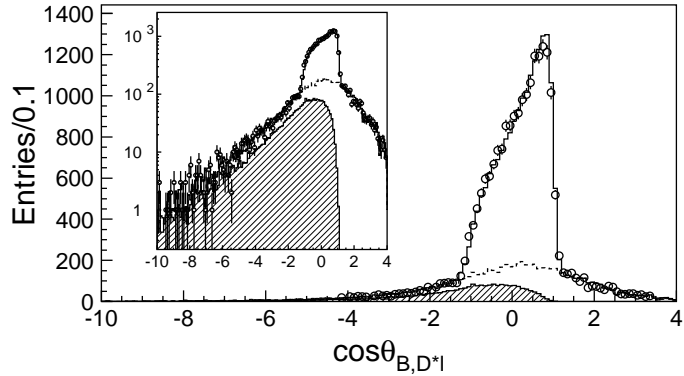}
  \end{center}
 \caption{The $\cosbdl$ distribution for the $\dslnu$ candidates. 
The circles with errors show
the data. The solid line is the fit result. The total background
and the $D^{**}\ell\nu$ component are shown by the dashed line
and the hatched area, respectively. The inset shows the same figure with
a logarithmic scale for the vertical axis.} 
\label{fig:cosb}
\end{figure}

%%%%%%%%%%%%%%%%%%
%\section{Event Selection}
%%%%%%%%%%%%%%%%%%
We use the decay chain
$B^{0} \to D^{*-} \ell^{+} \nu$, $D^{*-} \to \dzb \pi^-$, and
$\dzbkppm$, $\kppmpz$ or $\kppmpppm$~\cite{CC}.
The large branching fractions
and distinctive final states of the semileptonic decay 
allow for the efficient isolation of a high-purity $B^0$ sample.
The event selection criteria are almost the same as those for our previous $CP$
violation measurement~\cite{CP1_Belle}.
Charged particles are selected from tracks with associated SVD hits.
Track momenta for $\dzbkppmpppm$ decays are required to be larger than 0.2
GeV/$c$. 
Candidate $\pi^0 \to \gamma\gamma$ decays are pairs of photons
with energies greater than 0.08~GeV that have an invariant
mass within 0.011~GeV/$c^2$ of $m_{\pi^0}$ and 
a total momentum greater than 0.2~GeV/$c$.
We require the invariant mass within $0.013~{\rm
GeV}/c^2$ of $m_{D^0}$ for $\dzbkppm$ or $\kppmpppm$, and
$-0.037 < M_{\kppmpz} - m_{D^0} < +0.023$~GeV/$c^2$ for $\dzbkppmpz$.
The mass difference between the $D^{*-}$ and
$\dzb$ candidates, $\mdiff$, should be within 1 MeV/$c^2$ of the nominal value.
The cms angle between the $D^{*-}$ candidate
and a lepton (an electron or a muon that has
a cms momentum within $1.4<p^{\rm cms}_{\ell}<2.4$ GeV/$c$)
is required to be greater than 90 degrees.
The energies and momenta
of the $B$ meson and the $D^*\ell$ system in the cms should satisfy
$\mnu^2 = (E^{\rm cms}_B-E^{\rm cms}_{D^*\ell})^2-
|\vec{p}^{\rm ~cms}_B|^2-|\vec{p}^{\rm ~cms}_{D^*\ell}|^2+
2|\vec{p}^{\rm ~cms}_B|\,|\vec{p}^{\rm ~cms}_{D^*\ell}|\cosbdl$,
where $\mnu$ is the neutrino mass and $\thetabdl$ is the
angle between
$\vec{p}^{\rm ~cms}_B$ and $\vec{p}^{\rm ~cms}_{D^*\ell}$.
We calculate $\cosbdl$ setting $\mnu=0$.
Figure~\ref{fig:cosb} shows the $\cosbdl$ distribution.
The signal region is defined as $|\cosbdl|<1.1$.

%%%%%%%%%%%%%%%%%%%%
%%%%%%%%%%%%
%
We identify
the flavor of the accompanying $B$ meson from
the properties of the decay products~\cite{CP1_Belle}.
Several categories of well measured tracks
that have a charge correlated with the $b$ flavor are selected:
high momentum leptons
from  $b\to$ $c\ell^-\overline{\nu}$,
lower momentum leptons from  $c\to$ $s\ell^+\nu$,
charged kaons and $\Lambda$ baryons from $b\to$ $c\to$ $s$,
high momentum pions originating from decays of the type
$B^0\to$ $D^{(*)-}X$ (where $X = \pi^+, \rho^+$, $a_1^+, {\rm etc.})$, and
slow pions from $D^{*-}\to$ $\overline{D}{}^0\pi^-$.
Information extracted from each track is combined for the
$b$-flavor determination,
taking into account correlations
in case tracks in more than one category are present.
For each flavor decision,
we assign a MC-determined flavor-tagging dilution factor $r$,
which ranges from $r=0$ for no flavor
discrimination to $r=1$ for unambiguous flavor assignment.
It is used only to sort data into six intervals of $r$, according to
estimated flavor purity.
More than 99.5\% of the events are assigned a non-zero value of $r$.

We reconstruct the $\bzdslnu$ decay vertex using
the $\dzb$
trajectory, the lepton track and the
interaction-point profile (IP) convolved with the
finite $B$ flight length in the plane perpendicular to the
$z$ axis (21 $\mu$m).
The reduced $\chi^2$ of the vertex
is required to be less than 15.
The method of reconstructing the tagging side $B$ vertex is described
elsewhere~\cite{CP1_Belle}.

We find 16397 candidates
after flavor tagging and vertex reconstruction.
The signal fraction is estimated to be 80.4\%.
The backgrounds consist of fake $D^*$ mesons (7.8\%), 
$B \to D^{**}\ell\nu$ events (7.4\%), 
random combinations of
$D^*$ mesons with leptons with no angular correlation (2.6\%; called 
``uncorrelated background'') and continuum events (1.8\%).
Here $D^{**}$ consists of non-resonant $D^*\pi$ components and charmed mesons
heavier than $D^*$.
The background due to a combination of a fake lepton
and a true $D^*$ from the same $B$ meson is
estimated with MC to be negligible.
We estimate the fake $D^*$ background fraction from
the $\dzb$ mass sideband events and from fake $D^*$ events 
reconstructed with wrong-charge slow pions.
The uncorrelated background fraction is evaluated by
counting candidates where
we invert the lepton momentum vector artificially.
We estimate the continuum background fraction by scaling the off-resonance
data (2.3 fb$^{-1}$) with the integrated luminosity.
We fit the $\cosbdl$ distribution in a range 
$-10 < \cosbdl < 1.1$
to estimate the $B\rightarrow D^{**}\ell \nu$ background
fraction;
the $\cosbdl$ shapes for the signal and 
$B\rightarrow D^{**}\ell \nu$
are modelled using MC and
all the other background fractions and distributions
are fixed from the aforementioned special background
samples.

%%%%%%%%%%%%%%%%%%%%%%%
%\section{Resolution function} 
%%%%%%%%%%%%%%%%%%%%%%%
The $\dt$ resolution function for the signal, $R_{\rm sig}(\Delta t)$, 
is expressed as
\begin{eqnarray*}
 R_{\rm sig}(\Delta t) &=& g_1G(\Delta t; \mu_1,\sigma_1)+
  (1-g_1)G(\Delta t; \mu_2,\sigma_2), \\
 \sigma_{1(2)} &=& S_{1(2)}\sqrt{\sigma_{\rm rec}^2+\sigma_{\rm tag}^2},
\end{eqnarray*}
where $G(x;\mu_{1(2)}, \sigma_{1(2)})$ is the main (tail) Gaussian 
component with $\mu_{1(2)}$ and
$\sigma_{1(2)}$ as the mean and standard deviation, respectively, 
$g_1$ is the fraction of the main component,
$S_{1(2)}$ is a scale factor that corrects
our imperfect error estimation, and $\sigma_{\rm rec(tag)}$ is
the $\Delta t$ error
calculated from the vertex error for the reconstructed (tagged) $B$ meson
determined for each event.
We extract the above parameters
from the $\dt$ distribution of the candidate $\bzdslnu$ events
without distinguishing between different flavor assignments.
The signal probability density function (PDF) is given by
$F_{\rm sig}(\dt) = \int^{\infty}_{-\infty}
  \Lambda(\Delta t';\taub)
  R_{\rm sig}(\Delta t-\Delta t')d\Delta t'$,
where $\Lambda(\dt;\taub) = \exp(-|\dt|/\tau_{B^0})/(2\tau_{B^0})$.
We define the likelihood value for each event as
$L_i = (1-f_{\rm bg})F_{\rm sig}(\dt_i)+f_{\rm bg}F_{\rm bg}(\dt_i)$,
where $f_{\rm bg}$ is the overall background fraction of 0.196  
and $F_{\rm bg}(\dt_i)$ is the background PDF given by
$\sum_{k}f_kF_k(\dt_i)$. Here $F_k$ and $f_k$ are the PDF
and the fraction, respectively,
for each of the four background components.
We use the signal PDF for the $B \rightarrow D^{**} \ell \nu$ component.
For the other background components, we use
$F_k(\Delta t) = \int^{\infty}_{-\infty}
[(1-f_{\delta k})\Lambda(\Delta t';\tau_k)
+f_{\delta k}\delta(\Delta t')]G(\Delta t-\Delta t'; \mu_k, \sigma_k))d\Delta t'$, and
$\sigma_k = S_k\sqrt{\sigma_{\rm rec}^2+\sigma_{\rm tag}^2}$.
Here $\delta(\Delta t')$ is the Dirac's delta function that accounts for
components with small or zero lifetimes.
The parameters in $F_k(\dt)$
are obtained from 
the upper $\mdiff$ sideband ($0.155 < \mdiff < 0.165$~GeV/$c^2$)
for fake $D^*$,
the lepton-momentum-inverted events for uncorrelated background,
and off-resonance data for continuum background, respectively.
We perform a likelihood fit to determine $R_{\rm sig}(\dt)$
with $\tau_{B^0} = 1.548$ ps~\cite{PDG2000} and 
with the background parameters fixed to the obtained values.
We find that
the fraction of the main component is large ($g_1 =0.87^{+0.06}_{-0.09}$)
and the $\Delta t$ error estimation is correct
($S_1 = 0.99^{+0.09}_{-0.10}$). 
The typical rms resolution on $\Delta t$ is 1.43 ps.

%%%%%%%%%%%%%%%%%%%%%%%%
%\section{delta M_d fit}
%%%%%%%%%%%%%%%%%%%%%%%%
We determine $\dmd$
by an unbinned maximum-likelihood
fit to the $\dt$ distributions.
We define the likelihood value for each event as
follows:
\begin{eqnarray*}
L^{\rm OF(SF)}_i &=&
(1-f_{\rm bg}^l)\{(1-f_{D^{**}\ell\nu})F^{\rm OF(SF)}_{\rm sig}(\Delta t_i)\\
&+&f_{D^{**}\ell\nu}F^{\rm OF(SF)}_{D^{**}\ell\nu}(\Delta t_i)\}\\
&+& f_{\rm bg}^l\sum_{k}f_k^l f^{\rm OF(SF)}_{lk}F^{\rm OF(SF)}_k(\dt_i),
\end{eqnarray*}
where OF (SF) denotes $\bz\bzb$ ($\bz\bz$ or $\bzb\bzb$),
i.e. a state with
the opposite (same) flavor,
$F_{\rm sig}^{\rm OF(SF)}$,
$F^{\rm OF(SF)}_{D^{**}\ell\nu}$ and
$F^{\rm OF(SF)}_k$ are
PDFs for the OF (SF) signal events,
$D^{**}\ell\nu$ decays and other backgrounds, respectively,
$f_{\rm bg}^l$ ($l=1,6$) is
an overall background fraction excluding $B \to D^{**}\ell\nu$
in each $r$ region
and $(1-f_{\rm bg}^l)f_{D^{**}\ell\nu}$ corresponds to a 
fraction of $D^{**}\ell\nu$ decays.
Other background fractions
$f_{lk}^{\rm OF(SF)}$, where
the relation $f^{\rm OF}_{lk} + f^{\rm SF}_{lk} = 1$ holds,
are obtained from the control samples
for the uncorrelated and fake $D^*$ backgrounds,
and from MC events for the continuum.

The PDFs for the OF and SF signal events are given by
\begin{eqnarray*}
   F^{\rm OF(SF)}_{\rm sig}(\Delta t) &=& \int^{\infty}_{-\infty}
    {\cal P}^{\rm OF(SF)}_{\rm mix}(\Delta t')
    R_{\rm sig}(\Delta t-\Delta t')d\Delta t',
\end{eqnarray*}
where
\begin{eqnarray*}
    {\cal P}^{\rm OF(SF)}_{\rm mix}(\dt) =
    \frac{e^{-|\dt|/\taub}}{4\taub}
    \{1 \pm (1-2w_l)\cos (\dmd \dt)\}.
\end{eqnarray*}
Here wrong tag fractions $w_l$ are also determined simultaneously.
The PDF for $B \rightarrow D^{**}\ell\nu$ is given by a sum of $B^0$
and $B^{+}$ components as
$F^{\rm OF(SF)}_{D^{**}\ell\nu}(\Delta t) = (1-f_{B^{+}})F^{\rm OF(SF)}_{\rm sig}(\Delta t) +f_{B^{+}}F^{\rm OF(SF)}_{B^{+}}(\Delta t)$,
where $f_{B^+}$ is the $B^+$ fraction 
in the $B \rightarrow D^{**}\ell \nu$ background.
%MH We use the signal PDF
%MH for the $B^0 \rightarrow D^{**-}\ell^+\nu$ background.
The $B^+\rightarrow D^{**}\ell\nu$ background PDFs
$F^{\rm OF(SF)}_{B^+}(\Delta t)$
are given by
the following functions convolved with $R_{\rm sig}$:
${\cal P}^{\rm OF}_{B^+}=(1-w_{B^+}^l){\cal P}_{B^+}$ and
${\cal P}^{\rm SF}_{B^+}=w_{B^+}^l{\cal P}_{B^+}$,
where 
${\cal P}_{B^+}(\dt) = (1-r_{\delta}f_{\delta})\Lambda(\dt;r_{B^+}\tau'_{B^+})+ r_{\delta}f_{\delta}\delta(\dt)$ and
$w_{B^+}^l$  is the wrong tag fraction determined from
a $B^+ \to \dzb \pi^+$ sample.
Since the fake $D^*$ background
includes a mixing component, we use a function that has the same
form as the PDF for $B \to D^{**}\ell\nu$
with parameters obtained from the $\mdiff$ sideband events.
Other background PDFs do not distinguish between SF and OF events
and are the same as those used to determine the resolution function.

We perform the fit with 10 free parameters
(listed in Table~\ref{table:mixingresult})
to the $\Delta t$ distributions of SF and OF events in the signal
region and the $B\rightarrow D^{**}\ell\nu$ dominant region defined 
as $-10 < \cosbdl < -1.1$.
In this way, the background parameters 
$f_{B^+}$, $f_{\delta}$ and $\tau'_{B^+}$
are determined simultaneously.
Additional correction factors,
$r_{\delta}$ and $r_{B^+}$, are introduced only in the signal
region to account for the difference between the two regions.
We use MC to determine
$r_{\delta}=0.62^{+0.14}_{-0.12}$ and $r_{B^+}=1.04\pm 0.02$.

%%%%%%%%%
%\section{Result}
%%%%%%%%%

\begin{figure}
 \begin{center}
  \includegraphics[width=0.6\textwidth,clip]{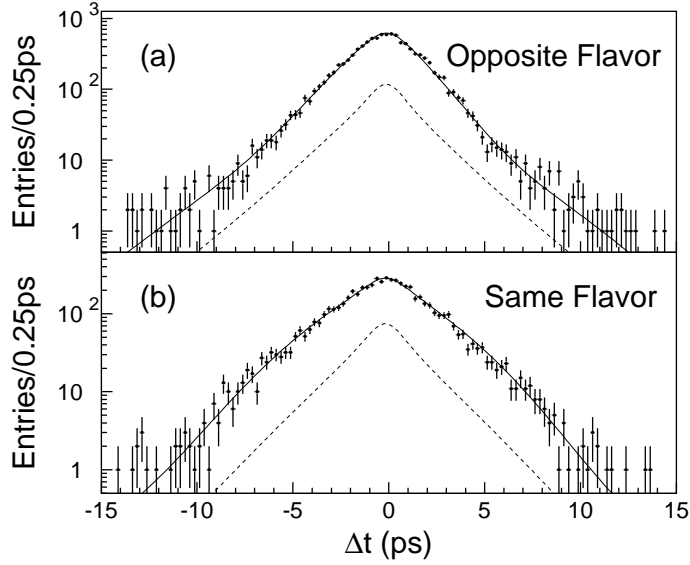}
  \caption{
  The $\dt$ distributions for
  (a) the OF events and (b) the SF events.
  The solid lines are the result of the unbinned maximum likelihood fit.
  The dashed lines show the background distribution.
  }
  \label{fig:deltat_wdm}
 \end{center}
\end{figure}

\begin{figure}
 \begin{center}
  \includegraphics[width=0.6\textwidth,clip]{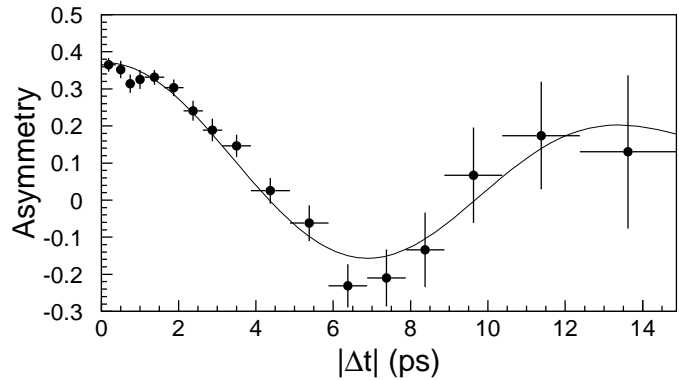}
  \caption{
  The observed time-dependent flavor asymmetry. The curve is the result of
  the unbinned maximum likelihood fit.
  }
  \label{fig:mixingasymmetry}
 \end{center}
\end{figure}

The fit result is summarized in Table~\ref{table:mixingresult}.
Figure~\ref{fig:deltat_wdm} shows the observed $\dt$ distributions
for the OF and SF events.
Figure~\ref{fig:mixingasymmetry} shows the corresponding flavor asymmetry,
${\cal A}(\dt)=[{\rm N}_{\rm OF}(\dt)-{\rm N}_{\rm SF}(\dt)]/[{\rm N}_{\rm OF}(\dt)+{\rm N}_{\rm SF}(\dt)]$,
where ${\rm N}_{\rm OF(SF)}$ denotes the number of OF (SF) events.
\begin{table}[hbtp]
  \caption{Summary of mixing fit. Errors are statistical only. For each wrong tag fraction,
  the $r$ interval and the number of candidate events are also shown.}
  \begin{ruledtabular}
  \begin{tabular}{ll}
  parameter & result \\
\hline
   $\dmd$ & $\DMDVAL\pm \DMDERRSTA$ ps$^{-1}$\\ 
   $w_1~(0 < r \le 0.25$, 6360 events) & $0.467\pm 0.010$ \\ 
   $w_2~(0.25 < r \le 0.5$, 2364 events) & $0.360\pm 0.016$ \\ 
   $w_3~(0.5 < r \le 0.625$, 1453 events) & $0.254\pm0.020$ \\
   $w_4~(0.625 < r \le 0.75$, 1702 events) & $0.182\pm0.017$ \\ 
   $w_5~(0.75 < r \le 0.875$, 1958 events) & $0.103\pm0.014$ \\ 
   $w_6~(0.875 < r \le 1$, 2560 events) & $0.032\pm0.010$ \\ 
   $f_{B^+}$ & $0.70 ^{+0.12}_{-0.13}$ \\ 
   $f_{\delta}$ & $0.28^{+0.10}_{-0.09}$ \\ 
   $\tau'_{B^+}$ & $1.87^{+0.24}_{-0.18}$ ps\\
  \end{tabular}
  \end{ruledtabular}
  \label{table:mixingresult}
\end{table}
%
%%%%%%%%%%%%%%
%%%%%%%%%%%%
%
%%%%%%%%%%%%%%
%%%%%%%%%%%%

%%%%%%%%%%%%%%%%%%%%
%\section{Systematic Errors}
%%%%%%%%%%%%%%%%%%%%
The systematic errors are summarized in Table~\ref{tbl:syserr}.
\begin{table}[hbtp]
  \caption{Summary of the systematic errors (ps$^{-1}$) on the $\Delta m_d$ measurement.}
  \begin{ruledtabular}
  \begin{tabular}{lr}
   source               & error (ps$^{-1}$)\\ \hline
   $D^{**}$ branching fractions &   0.007 \\
   $|\Delta t|$ range   &   0.007 \\ 
   Background shape     &   0.006 \\
   Resolution function  &   0.006 \\
   $B^0$ lifetime       &   0.005 \\
   Fit bias             &   0.004 \\ 
   Background fraction  &   0.003 \\
   $B\rightarrow D^{**}
   \ell\nu$ fraction    &   0.002 \\ 
   IP constraint        &   0.002 \\
   $B^{\pm}$ wrong 
   tag fraction         & $<0.001$ \\
   $B^{\pm}$ shape 
   parameter            & $<0.001$ \\
\hline
   total                & 0.015 \\
  \end{tabular} 
  \end{ruledtabular}
  \label{tbl:syserr}
\end{table}
%
%\subsubsection*{$D^{**}$ branching fractions}
$D^{**}$ branching fractions
used in this analysis are based on
theoretical assumptions~\cite{dsslnu}.
We set each such branching fraction to unity in the MC
(with all others set to zero), and repeat the analysis;
we take the largest variation on the $\dmd$ result 
as the systematic error.
%
%\subsubsection*{$|\Delta t|$ cut}
To account for uncertainties in the tails of the vertex resolution,
we measure $\dmd$ by setting the upper limit on $|\dt|$
that ranges from 5 to 55 ps. 
We take the largest difference from the main result,
which is obtained without the $|\dt|$ upper limit,
as a systematic error.
Systematic errors from the background PDFs
are obtained by varying each shape parameter individually,
repeating the fit procedure, and adding 
each contribution in quadrature.
We also perform a MC study where we obtain
background PDFs by two methods: one from the background
control samples, and the other
directly from the signal region. 
The difference between two $\Delta m_d$ fit results 
is included in the systematic error.
%
%\subsubsection*{Fit bias}
A fit with the $B \rightarrow D^*\ell\nu$ and 
$B \rightarrow D^{**}\ell\nu$ MC events yields the $\dmd$ value that is
consistent with the input value within 1.7$\sigma$. We conservatively take the
difference, which we attribute to MC statistics, 
as a systematic uncertainty.
The systematic error due to the IP constraint is estimated by
varying ($\pm10 \mu$m) the smearing used to account for the
$B$ flight length.
Other sources of systematic errors are obtained
by changing each parameter by 1$\sigma$,
repeating the fit procedure and adding each contribution in quadrature.
%
%%%%%%%%%%%%%%%%%%%%%
%\section{Consistency checks}
%%%%%%%%%%%%%%%%%%%%%
%\subsubsection*{The $r$ dependence}
We also perform
a $\Delta m_d$ fit in each $r$ region and find no systematic trend.
%
%%%%%%%%%%%%%
%\section{Conclusion}
%%%%%%%%%%%%%

In summary,
we have measured the $\bz$-$\bzb$ mixing parameter $\Delta m_d$ using 
$B^0 \rightarrow D^{*-}\ell^+\nu$ decays in a
29.1~fb$^{-1}$ data sample 
collected  with
the Belle detector at the KEKB
$e^+e^-$ collider
operating at the $\Upsilon(4S)$ resonance.
From an unbinned maximum likelihood fit to
the $\dt$ distributions for
$B$ pairs with the same and opposite flavors,
we obtain 
\begin{eqnarray*}
 \dmd = \DMDRESULT.
\end{eqnarray*}
The result is one of the most precise measurements performed so far,
and is consistent with 
the world average value of $\Delta m_d = 0.472 \pm
0.017$ (ps$^{-1}$)~\cite{PDG2000}
as well as other recent measurements~\cite{OTHERDMD}.

%%%%%%%%%%%%%%%%%%
%\section{Acknowledgments}
%%%%%%%%%%%%%%%%%%
We wish to thank the KEKB accelerator group for the excellent
operation of the KEKB accelerator.
We acknowledge support from the Ministry of Education,
Culture, Sports, Science, and Technology of Japan
and the Japan Society for the Promotion of Science;
the Australian Research Council
and the Australian Department of Industry, Science and Resources;
the National Science Foundation of China under contract No.~10175071;
the Department of Science and Technology of India;
the BK21 program of the Ministry of Education of Korea
and the CHEP SRC program of the Korea Science and Engineering Foundation;
the Polish State Committee for Scientific Research
under contract No.~2P03B 17017;
the Ministry of Science and Technology of the Russian Federation;
the Ministry of Education, Science and Sport of the Republic of Slovenia;
the National Science Council and the Ministry of Education of Taiwan;
and the U.S.\ Department of Energy.


\begin{thebibliography}{99}
\bibitem{CC}
Throughout this Letter, 
the inclusion of the charge conjugate mode decay is implied.
\end{thebibliography}
\end{document}